\documentclass{raa_rb}
\usepackage[pagebackref=true]{hyperref}
\usepackage{graphicx,times}
\usepackage{natbib}
\usepackage{amssymb,amsmath}
\usepackage{newtxtext,newtxmath}
\usepackage{booktabs}
\usepackage[T1]{fontenc}
\usepackage{comment}

\bibpunct{(}{)}{;}{a}{}{,}
\usepackage{xcolor}
\bibpunct{(}{)}{;}{a}{}{,}

\usepackage{soul}
\setstcolor{red}
\definecolor{dkblue}{RGB}{54, 86, 169}

\newcommand{\newuppi}{\mathrm{\pi}}
\newcommand{\e}{\mathrm{e}}

\newcommand{\C}{\textsf{\textbf{C}}}

\newcommand{\BF}{\mathcal{B}}

\newcommand{\tr}{\textsf{T}}
\newcommand{\toa}{\boldsymbol{t}}
\newcommand{\res}{\delta\boldsymbol{t}}

\newcommand{\ttwo}{{\tt TEMPO2}}
\newcommand{\tn}{{\tt TEMPONEST}}

\newcommand{\psrx}{{\tt PSRCHIVE}}
\newcommand{\dspsr}{{\tt DSPSR}}
\newcommand{\ftwo}{{\tt FORTYTWO}}

\newcommand{\psr}{PSR~J1713+0747}
\usepackage{lineno}

\begin{document}
\title{Chinese Pulsar Timing Array upper limits on microhertz gravitational waves from supermassive black-hole binaries using PSR~J1713+0747 FAST data}

 \volnopage{ {\bf 20XX} Vol.\ {\bf X} No. {\bf XX}, 000--000}
   \setcounter{page}{1}

 \author{R. Nicolas Caballero
	 \inst{*1,2}, 
  Heng Xu\inst{*1}, 
  Kejia Lee\inst{*3,1,4}, 
  Siyuan Chen\inst{5,2,3}
  Yanjun Guo\inst{6},
  Jinchen Jiang\inst{1},
  Bojun Wang\inst{1},
  Jiangwei Xu\inst{3,1,2},
  Zihan Xue\inst{3,1,2}
  }
  
	 \institute{  National Astronomical Observatory, Chinese Academy of Sciences, 
	 Beijing 100101, P.~R.~China 
\and
Kavli Institute for Astronomy and Astrophysics, Peking University, 
	 Beijing 100871, P.~R.~China; {\it caballero.astro@gmail.com}
  \and
	 Department of Astronomy, School of Physics, Peking University, Beijing 100871, P.~R.~China; \\
  \and
	 Beijing Laser Acceleration Innovation Center, Huairou, Beijing 101400, P.~R.~China
  \and
  Shanghai Astronomical Observatory, Chinese Academy of Sciences, 80 Nandan Road, Shanghai 200030, P.~R.~China
\and
Max-Planck-Institut für Radioastronomie, Auf dem Hügel 69, 53121 Bonn, Germany 
	 \\
  {\small Received 20XX Month Day; accepted 20XX Month Day}
}

\abstract{We derive the gravitational-wave (GW) strain upper limits from resolvable supermassive black-hole binaries using the data from the Five-hundred-meter Aperture Spherical radio Telescope (FAST), 
in the context of the 
Chinese Pulsar Timing Array project. We focus on circular orbits in
the $\mu$Hz GW frequency band between $10^{-7}$ and $3\times10^{-6}$~Hz. This frequency band is higher than the traditional pulsar timing array band and is less explored. We used the data of the millisecond pulsar 
PSR~J1713+5307 observed between August 2019 and April 2021. A dense observation campaign was carried out in September 2020 to allow for the $\mu$Hz band coverage. Our sky-average continuous source upper limit at the 95\% confidence level 
at 1$\mu$Hz is 1.26$\times10^{-12}$, while the 
same limit in the direction of the pulsar is 4.77$\times10^{-13}$. 
}

\authorrunning{R.~N.~Caballero et al. }            
\titlerunning{CPTA microhertz GW limits from PSR J1713+0747}  
   \maketitle
\section{Introduction}

The primary work for the search of gravitational waves (GWs) 
at nHz frequencies over the last decades has been conducted 
using radio pulsars. In particular, the work is based
on the employment of Pulsar Timing Arrays \citep[PTAs;][]{fb1990}, 
which are ensembles of millisecond pulsars (MSPs) at different sky locations.
MSPs are chosen for the work of GW searches 
due to their remarkably stable rotations \citep[see e.g.][]{vbc+2009}.
At the same time, they allow for
the highest-precision recording of pulse arrival times (times-of-arrival; ToAs), 
which can be of the order of tens of nanoseconds, as is the case with data in this 
paper.
By comparison, the canonical pulsar population which comprises of
the younger pulsars, exhibits a substantial amount of
rotational irregularities \citep{hlk2010,psj+201}. These manifest in the form of
regular glitches, nulling and mode changes, and in strong stochastic irregularities 
known as timing noise (also referred to as red noise and spin noise),
which reduce the sensitivity to GW signals \citep[see e.g.][]{cll+2016}.
On the other hand, MSPs have very smooth and regular rotations
which result in very small levels of stochastic timing noise \citep[e.g.][]{vbc+2009,cll+2016,lsc+2016}, and only a couple of glitches 
have been previously observed in MSPs \citep{cb2004,mjs+2016}.
Profile changes in MSPs are generally subtle,
in the form of profile instabilities,
and can be addressed in standard timing methods
\citep[e.g.][]{kxc1999,lkl2015,slk2016,krh+2020}.
Profile changes that are severe and affect significantly
the long-term timing of MSPs are rare,
but \psr{} has recently exhibited such a change which 
led to the limitation
in the total dataset used in this study (see Section~\ref{sec:data}).

PTAs are mostly sensitive in the nHz part of the GW spectrum, 
where the primary goal is to observe GWs emitted by supermassive black-hole binaries (SMBHBs). 
Apart from GWs from resolvable SMBHBs (see e.g. \citealt{tho1989,jll+2004,sv2010,lwk+2011}), 
PTAs are also sensitive to stochastic 
GW background (GWB) signals.  Such stochastic signals may be the result
of the superposition of GWs 
from the cosmic population of SMBHBs \citep{rr1995,jb2003}, 
cosmic strings \citep{kib1976,sbs2012},
and relic GWB from quantum fluctuations in the Early Universe, 
especially from the inflationary era \citep{kib1976}. 
Recently, the CPTA and other PTA groups, have reported 
statistical evidence for the existence of a nHz GWB \citep{xcg+2023,epta2023,ng2023,ppta2023}, 
each group
PTAs have not yet detected a single GW source.

The observational GW frequency window for PTAs 
is determined by the total timespan of the
data set, $T$, and the observational cadence, $\Delta{}t$. 
The upper frequency limit is determined by the Nyquist theorem 
and is $f_{\textrm{max}}\approx1/(2\Delta{}t)$ for regular sampling. 
For irregular sampling, 
which is the case for real pulsar timing data, the upper frequency limit can, in principle, be much higher than the `traditional' Nyquist frequency (based on arguments similar to that of \citet{2006MNRAS.371.1390K}). In practical PTA GW detection, on the other hand, the interesting upper frequency limit is mainly determined by the data sensitivity. 
The lower recoverable frequency is $f_{\textrm{min}}\approx1/T$, 
as any power in the pulsar ToAs below this frequency is 
effectively absorbed by the quadratic term of the 
pulsar's rotational-frequency derivative \citep[see][]{lbj+2012}. 
This term is always present
in timing models of radio pulsars, and accounts for the slowing-down
of the pulsar rotation as it loses rotational energy through the emitted radiation.

Typical pulsar data sets have decadal time spans for a large number of pulsars,
and have the highest GW sensitivity around the lowest frequencies of the spectrum.
Single-telescope data seldom can have cadence below $\sim$1 week, 
but multi-telescope data combination can achieve daily cadence.
In this way, PTAs can probe single source GWs (SSGWs) up to the $\mu$Hz band.
So far, the $\mu$Hz regime for SSGW from SMBHBs with pulsar timing
has been explored using single-pulsar data \citep{yss+2014,dlc+2014,psb+18}.
As a consequence, these studies can only provide GW amplitude strain upper limits,
as GW detection requires observations by multiple pulsars.
The current paper also places upper limit using single-pulsar data.

\psr{} is one of the most observed
and thoroughly studied MSPs.
Due to its sky location, that permits observations from 
both the north and south hemisphere, its long-term stability
and high brightness that results in high ToA measurement precision,
\psr{} has always been a primary PTA target
for GW search efforts.
It has been continuously observed by the three founding PTAs and has also been the subject of the an IPTA 24-hour global 
observing campaign \citep{dlc+2014}, with nine participating telescope
observing the source as it became visible in their local sky,
allowing rare investigations of noise properties
on intermediate timescales. 
In this paper, we use \psr{} data from the Five-hundred-meter Aperture Spherical radio Telescope \citep[FAST;][]{Jiang19SCPMA} in China to place
new and independent upper limits for the stain amplitude of
SSGWs from circular SMBHBs, in the $\mu$Hz regime.

The rest of the paper is organized as follows:
in section~\ref{sec:data}, we describe the data used for the study.
In section~\ref{sec:spnta} we present the 
pulsar timing model and the data's noise properties, while
in section~\ref{sec:gws} we derive the SSGW strain upper limits.
Finally, we discuss our conclusions in section~\ref{sec:conclude}.

\section{Data}
\label{sec:data}
PSR~J1713+0747 was observed using the 19-beam receiver of the FAST radio telescope from August 2019 to April 2021 with a cadence of about 1 week, and we conducted a dense observation campaign with nearly 2-day cadence in September of 2020. Since this pulsar was found to show a profile changing event \citep{Xu2021Atel}, the data after 2021 April 16th are not included in this work. The receiver centers at 1.25\,GHz with a bandwidth of 500\,MHz. We used the ROACH2 system to record the data in search mode with a 49.152 ${\rm \mu s}$ sampling time and 4094 channel mode. 
Each observation was typically 20-30 minutes, although some were longer, 
up to 3 hours. In total, we include 56.2 hours of observations. The data were off-line folded every 30 seconds using the standard pulsar data reduction software \dspsr{} \citep{dspsr}, and the polarization calibration was performed with the pulsar analysis software package \psrx{} \citep{psrchive} using the single-axis model. Following polarization calibration, any data corrupted by radio frequency interference (RFI) was removed manually. The clean data were finally time-integrated every 15 or 20 mins and the whole band was split
into 16 subbands to allow for the measurement and corrections of frequency dependent noise, e.g. dispersion measure (DM) variation.
The subband ToAs were generated following the Fourier-domain algorithm \citep{Taylor1992}  with the \textsc{pat} routine in \psrx{}. ToAs with signal-to-noise ratio (S/N) lower than 8 were removed from the subsequent analysis.  
In total, 3436 ToAs from 65 observation session were used in this paper.

\begin{figure}
\includegraphics[width=\columnwidth]{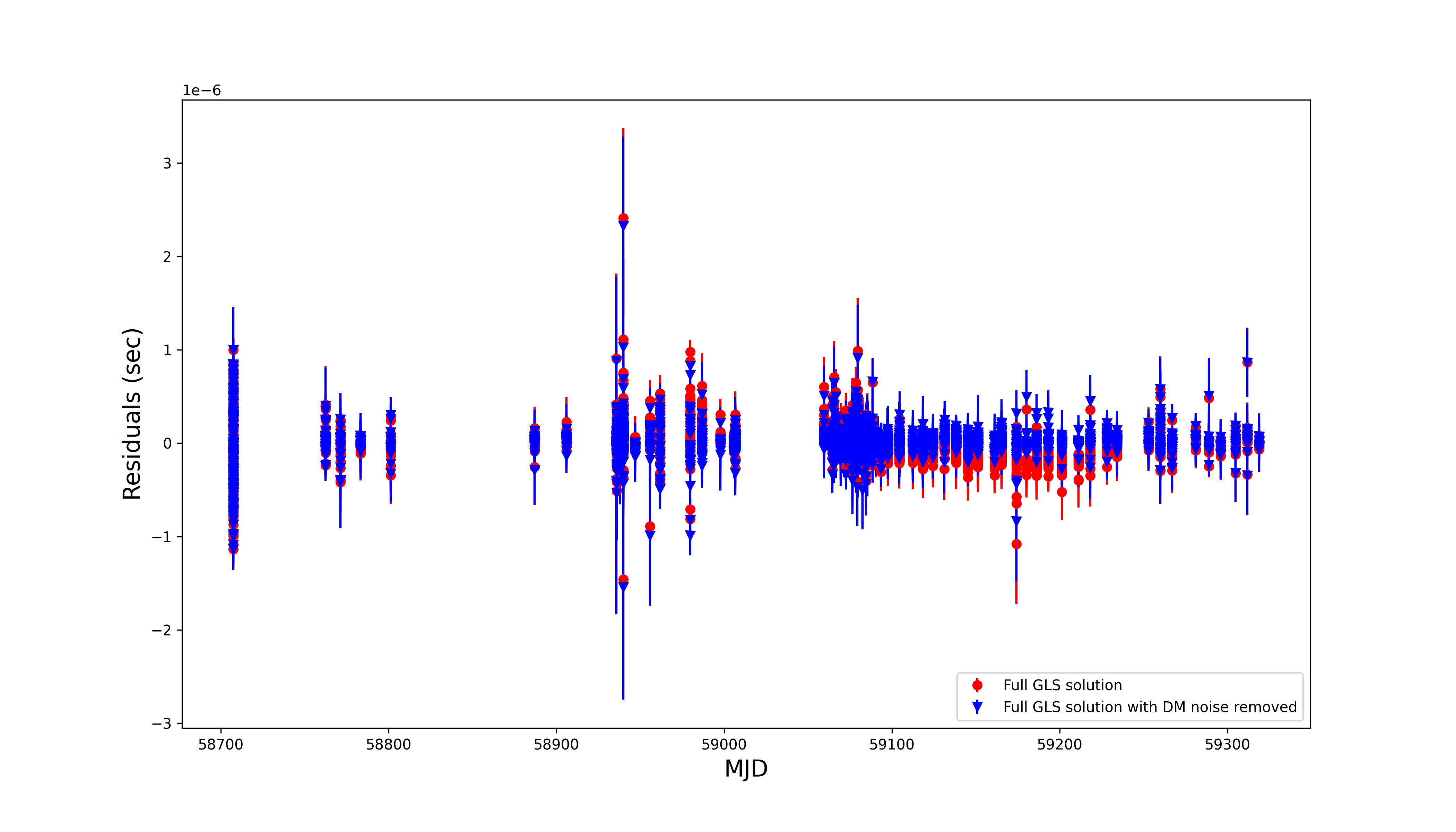}
\caption{Timing residuals (red circles) of \psr{} after
a generalized-least-squares fit of the timing model with \ttwo{}.
The covariance matrix is constructed based on the maximum-likelihood
values of the inferred noise parameters.
We also overplot (blue triangles) 
the same residuals after subtracting the stochastic noise due to DM variations.
The weighted root-mean-square of the two timing residuals are 115 and 93\,ns,
respectively.}
\label{fig:GLS_res}
\end{figure}

\section{Pulsar Timing Model and Noise Properties}
\label{sec:spnta}

The initial timing analysis of the data were performed using \ttwo{} \citep{hem2006},
providing a phase-coherent timing solution for the data. 
As \ttwo{} uses a linearized model and the solution 
is calculated via least-squares fit, the presence of any
red noise can bias the estimated values and uncertainties of the 
timing parameters \citep{chc+2011,vl2013}. 
We followed a Bayesian approach using \tn{} \citep{lah+2014} for the noise analysis,
searching for noise components that are standard in pulsar timing
\citep[see for example][and see further details in the next section]{lsc+2016}.
We used Bayesian model selection using Bayesian factors to 
determine how many components are supported by the data.
Detailed timing and noise analysis are presented 
for \psr{} and all MSPs timed by the CPTA
in separate papers (Xu et al., in prep.; Chen et al., in prep.).
Fig.~\ref{fig:GLS_res} shows the timing residuals from 
a generalized-least-squares fit of the timing model
including the noise model,
as well as the same residuals after subtracting the 
stochastic effects of the interstellar medium (see next section).
The latter residuals have a root-mean-square of only 93~ns.

\subsection{Noise model}
\label{sec:noise_model}
In general, for a one dimensional time series, 
noise can be divided in time-uncorrelated and time-correlated noise.
The former are referred to as white noise parameters and they include:
(i) a corrective multiplying factor on the ToAs uncertainties calculated
during the ToA estimation \citep{tw1982} named EFAC, 
(ii) a factor added in quadrature to the ToA uncertainties\citep{em1968,sc2010}, named EQUAD,
which tracks effects of pulse phase jitter, and
(iii) a parameter named ECORR \citep{abb2016}, which models the pulse jitter effects that are specifically correlated in subband ToAs across the observing bandwidth.
The interested reader can read more details in the CPTA pulsar-noise paper (Chen et al., in prep).

The time-correlated noise components included are:
(i) the red (achromatic) noise, stochastic noise pressumably associated with pulsar spin irregularities \citep{sc2010} and
(ii) stochastic DM variations \citep[see][]{yhc+2007a}.
DM is the column density of free electrons between the pulsar and Earth,
causing a delay $\delta\tau_{\rm{dm}}$ in the radio signal proportional 
to the inverse square of the observing frequencies, $\nu$
(i.e. $\delta\tau_{\rm{dm}}\propto\nu^{-2}$),
following the dispersion law of cold plasma \citep[e.g.][]{ll1960}.
These types of stochastic, correlated noise are typically
assumed to be wide-sense stationary signals, and are
modelled in the time-frequency domain \citep{lah+2013}.
Following the literature, 
we use a power-law spectrum for red noise of the form,
\begin{equation}
S(f) \propto A^2f^{-\gamma}\,,
\label{eq:spectrum}
\end{equation}
where $S(f)$ is the power spectral density, $A$ is the spectrum's amplitude and $\gamma$ is the spectral index. 

\subsection{Bayesian parameter estimation and model selection}
\label{sec:bayes_theory}

We use Bayesian inference to perform parameter estimation, where
the model selection is based on the Bayes evidence comparasion
as previously described \citep{lsc+2016, cbp+2022}. We employed the software package \tn{} to perform the Bayesian noise modeling. Here, we give a brief mathematical description of the process,
which is also (partially) applied in the calculations of the 
SSGW amplitude limits discussed in the next section.

Bayes' theorem relates the posterior distribution (i.e. the `distribution' of inferred parameters, given the data) and the likelihood (the `distribution'  of data, given the model), as
\begin{equation}
\label{eq:bayesT}
P(\lambda|\mathcal{D})=\frac{L(\mathcal{D}|\lambda)p(\lambda)}{Z}\,,
\end{equation}
where
$\mathcal{D}$ denotes our data,
$H$ denotes the hypothesis (i.e. the model),
and $\lambda$ denotes the model parameters.
$P$ is the
posterior distribution,
$L$ is the likelihood function,
$p$ is the prior distribution,
and $Z$ is the Baysian evidence.

The likelihood function of pulsar timing was assumed to be Gaussian \citep{vlm+2009}, i.e.
\begin{equation}
\label{eq:psrliktoa}
L=\frac{\e^{-\frac{1}{2}(\toa-\toa_{\textrm{tm}})^{\tr}\C^{-1}(\toa-\toa_{\textrm{tm}})}}{\sqrt{(2\newuppi)^n|\C|}}\,,
\end{equation}
where $\toa$ is the recorded ToA, 
$\toa_{\textrm{tm}}$ is the ToA predicted by the timing model,
($\toa-\toa_{\textrm{tm}}$ is, therefore, equal to the timing residuals)
$\C$ is the total noise covariance matrix,
and $n$ is the number of ToA in the data.

The prior distribution represents our prior knowledge
or belief for the distribution of the unknown parameter.
This choice therefore can add information to the parameter
estimation and can lead to erroneous results if our belief of the
parameter distribution is informative but wrong. 
Theoretically, \emph{Jeffreys prior} represents a choice of prior being the least informative.
However its formal calculation can be difficult and approximations are
used. For example, for scale invariant parameters, such as the 
amplitude of the power-law red noise spectrum,
we use a prior uniform (flat) in the $\log$ space 
as an uninformative prior \citep{gre2005},
while one would use a prior uniform in linear space
to estimate conservative upper limits\footnote{By taking the `log', the physical units (e.g. Volts) of signal amplitude becomes dimensionless, and one removes the information inherited in the units system, i.e. the scale. Taking the uniform prior, one `believes' in higher value and assigns a higer probability to it, 
which aids in calculating \emph{conservative} upper bounds.}. 

The evidence $Z$ is accurately estimated when we want 
to perform model comparison and selection in order to 
decide which model better fits the data.
The evidence is the integrated likelihood over the
prior, and for $N$ parameters is defined as,
\begin{equation}
Z=\int L(\mathcal{D}|\lambda)p(\lambda)\,d^{N}\lambda.
\end{equation}
Larger $Z$ values denote more favourable models.
As a metric to choose whether a more complicated model
is required, $Z$ should be larger than a threshold.
We can measure this using the posterior odds ratio 
for two models, e.g. 0 and 1,
\begin{equation}
\label{eq:postodds}
R = \frac{Z_1}{Z_0}\frac{p_1}{p_0} = \BF_{10}\frac{p_1}{p_0} \,.
\end{equation}
where the ratio of the evidences $\BF_{10}$ is called the
Bayes factor. As the prior distributions for all models
we discuss in model selection analysis in this paper 
are the same, the Bayes factor is equal to the posterior odds ratio.
The distribution function of Bayes factor depends on the prior and likelihood. As a rule of thumb, a threshold of $\log_{10}(\BF_{10})>2$ is taken in the current paper to choose the preferred model, following \cite{kr1995}. 

\subsection{Noise analysis}
\label{sec:noise_analysis}

In the noise analysis,
we used the priors shown in 
Table~\ref{tab:spna_prios}, and 
employed \tn{}'s \emph{importance nested sampling} algorithm 
to perform posterior and evidence computation \citep[see][]{ski2004,fh2008}.
Our model selection process uses Bayes factors to 
select our preferred noise model from a set of nested models,
where the simplest models includes EFAC only.
We proceed trying the models with white-noise parameters
only as, EFAC+EQUAD, EFAC+ECORR, EFAC+EQUAD+ECORR.
The analysis results suggested that all three terms were required.
To test whether the data support red and/or DM stochastic noise,
we first assume EFAC+RN and EFAC+DM models 
with increasing number of frequency bins, to find the optimal
number of frequency bins for the RN and DM power-law spectrum models
using Bayes factors evaluations.
Once that number was defined we tested the EFAC+RN, EFAC+DM and EFAC+RN+DM models with the respective optimal numbers of frequency bins.
The analysis suggested that the data are sufficiently modelled 
using white noise+DM noise, and that there is not sufficient support
to add the RN noise component. We note that this does not mean that
a RN component, including a GWB, is not present. It means that if present, 
the current dataset is not sufficient to detect it with such
high statistical significance, and therefore model selection preferes the
simpler model that doesn't include it.
This is expected for a GWB from GW-driven SMBHBs whose stochastic
signal has a spectral index of 4.33, as a longer dataspan 
would be required to detect it: see for example \cite{jhl+2005} 
where (simulated) data similar to CPTA data 
require $\gtrsim$~3 years for such a GWB detection,
depending on the exact pulsar number and noise properties,
and \cite{xcg+2023} for the case of CPTA data.
The detected DM noise in the current data is a shallow power-law with
spectral index of $\approx$1.7 (see Table~\ref{tab:spna_res}), thus allowing its detection at the higher frequencies
where our data is more sensitive.
We confirmed that from all the white-noise
models with DM noise added, the most supported model 
is the EFAC+EQUAD+ECORR+DM model. We then confirmed that changing the 
number of frequency bins for the RN does not improve the support
to add the RN in the EFAC+EQUAD+ECORR+DM model, and that the optimal
number of DM noise frequency bins (43) remains the same with the optimal model
as in the case of the simpler EFAC+DM model.
Thus, the DM power-law signal spans in the frequency spectrum 
from 1/$T$ (1.9$\times$10$^{-8}$\,Hz) to the highest frequency of 1/43\,days$^{-1}$ (2.7$\times$10$^{-7}$\,Hz).

\begin{figure}
\includegraphics[width=\columnwidth]{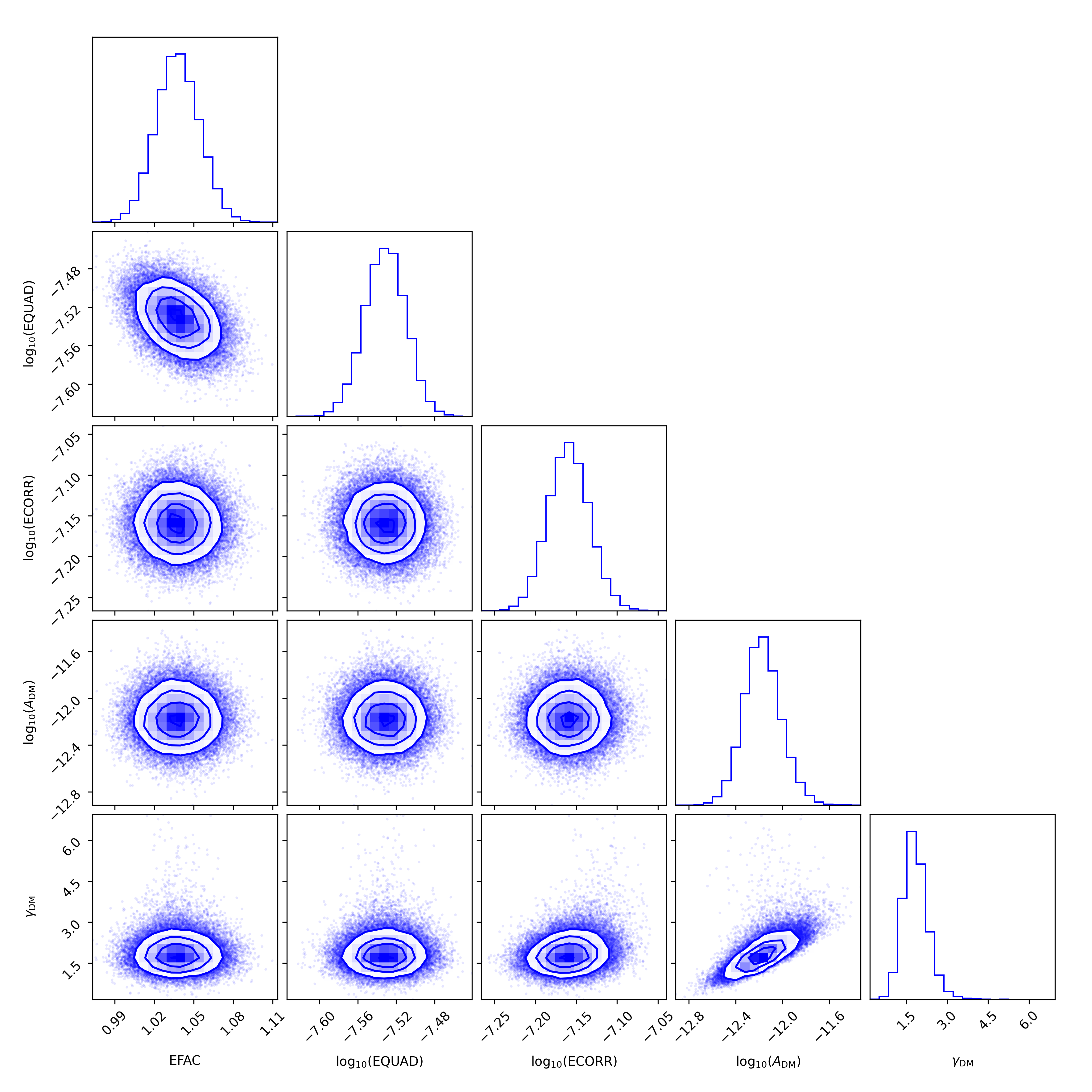}
\caption{Corner plot for the posterior distributions of the
single-pulsar noise analysis for the optimal noise model (see main text for details), displaying 2-D and 1-D marginalized distributions.}
\label{fig:spna_corner}
\end{figure}

Here, we present the results from the optimal model's noise analysis.
Fig.~\ref{fig:spna_corner} is the corner plot showing the 2-D and marginalized 1-D histograms
of the noise parameters. We note that while EFAC and EQUAD appear correlated,
ECORR does not appear correlate to either of EFAC and EQUAD. On the other hand,
all three of these parameter show correlation with the spectral index of DM.
This happens due to the very shallow DM power-law, i.e. because the
DM signal has significant power in higher frequencies. 
Table~\ref{tab:spna_res} gives the median and maximum-likelihood
values of the noise-parameters posterior distributions.

\begin{table} 
    \centering
    \caption{Ranges and types for the priors used in the 
    single-pulsar noise analysis of \psr{}. The prior types 
    Uni and log-Uni correspond to uniform priors in linear and logarithmic 
    space, respectively.}
    \label{tab:spna_prios}
    \begin{tabular}{ccc}
    \toprule
      Parameter & Range & type \\
            \midrule             
         EFAC & [0.1,5] & Uni\\
        $\log_{10}$(EQUAD) & [$-9$,$-5$] & log-Uni\\
        $\log_{10}$(ECORR) & [$-9$,$-5$] & log-Uni \\
        $\log_{10}$(A$_{\textrm{DM}})$ & [-20,-8] & log-Uni\\
        $\gamma_{\textrm{DM}}$ & [0,7] & Uni\\
        $\log_{10}$(A$_{\textrm{RN}})$ & [-20,-8] & log-Uni\\
        $\gamma_{\textrm{RN}}$ & [0,7] & Uni\\
    \bottomrule
    \end{tabular}
\end{table}

\renewcommand{\arraystretch}{1.1}
\begin{table} 
    \centering
    \caption{Overview of the results from the
    single-pulsar noise analysis of \psr{}, for the optimal noise 
    model (see main text for details). The table lists the median values of the
    1-D marginalized posterior distribution for each parameter, with the (asymmetric) 
    68\% uncertainties, as well as the maximum likelihood (ML) values.}
    \label{tab:spna_res}
    \begin{tabular}{ccc}
    \toprule
      Parameter & Median & ML \\
            \midrule             
        EFAC & 1.037$^{+0.017}_{-0.016}$ & 1.039\\
        $\log_{10}$(EQUAD) & -7.531$^{+0.021}_{-0.022}$ & -7.533 \\
        $\log_{10}$(ECORR) & -7.159$^{+0.0255}_{-0.0246}$ & -7.162 \\
        $\log_{10}$(A$_{\textrm{DM}})$ & -12.176$^{+0.163}_{-0.150}$ & -12.196\\
        $\gamma_{\textrm{DM}}$ & 1.804$^{+0.472}_{-0.395}$ & 1.678\\
    \bottomrule
    \end{tabular}
\end{table}

\section{Continuous gravitational-wave strain upper limits}
\label{sec:gws}

In the context of PTAs, the problem of SMBHB GW detection has been discussed extensively \citep[e.g.][]{jll+2004,lwk+2011,ses2013}.
In general, GWs from a SMBHB have two terms, which we refer to
as the '\emph{Earth term}' and the '\emph{pulsar term}', 
which quantify the spacetime distortion of the passing GW
at the Earth's and at the pulsar's vicinities, respectively.
The two-component SSGW signal is a direct result of the
large (thousands of years) 
time-delay between the pulsar and the Earth.
In this work, we focus on the case of monochromatic GWs and neglect the evolution of GW source.
We adopt the mathematical framework described in \cite{lwk+2011}.
The monochromatic GW is then described by seven parameters, 
namely the GW strain amplitude ($A$), 
the two GW-source position sky co-ordinates (RA,DEC), 
the GW angular frequency ($\omega_{\rm gw}$), 
the orbital inclination angle ($\iota$), 
the orientation angle of ascending node ($\psi$), 
and the GW initial phase 
($\phi_0$) at reference epoch $t_0$.
The term `monochromatic' refers to signals where we have
in effect the condition $\delta{}f<<\omega_{\rm gw}/2\pi$,
where $\delta{}f$ is the difference between the frequency of the Earth and pulsar terms.

As only one pulsar is considered,  our SSGW analysis is limited to the derivation
of robust upper limits for the GW strain amplitude.
We have employed two methods to estimate the upper limits.
We use 1) analytic calculations based on the Cram\'er-Rao
lower bound (CRLB) and 2) Bayesian inference analysis. In contrast to the Bayesian inference,
the CRLB calculations require
less computational resources and
are used to cross-check the Bayesian result, as the CRLB 
provides the theoretically best 
possible result \cite{fis63} and the Bayesian-inference
result must be `worse' than the CRLB in any case \citep[see discussion in][]{lwk+2011,cll+2016}.
We are interested in two results: (i) the sky-averaged and (ii) the best-sky.
Because we only use on pulsar, our analysis is insensitive to
the sky-direction opposite to the pulsar position 
(henceforth anti-pulsar direction), while it is optimal
around the direction of the pulsar.
For this reason, when computing the all-sky upper limit,
we exclude an area around the anti-pulsar direction (see Section~\ref{sec:bayeslims}).
All SSGW analysis was performed using 
the PTA data analysis package \ftwo{}\footnote{\url{https://github.com/caballero-astro/fortytwo}}
\citep{cll+2016}
which performs pulsar-timing noise and GW analysis.

\subsection{Cram\'er-Rao Lower-bound}
\label{sec:crlb}

The calculation of GW sensitivity curves for PTA data
using the CRLB was previously described in \cite{2016ASPC..502...19L,cll+2016}.
The CRLB is defined as the inversion of the Fisher information matrix, $\boldsymbol{\mathcal{I}}$.
Given the likelihood function, $L(\boldsymbol{\lambda}, {\bf x})$,
where ${\bf x}$ is the 
data and $\boldsymbol \lambda$ are the model parameters, the CRLB is described by:

\begin{equation}
\label{eq:CRLB1}
\textrm{Cov}(\boldsymbol{\lambda})=\langle 
\sigma_{\lambda_i}\sigma_{\lambda_j}\rangle \geq 
\boldsymbol{\mathcal{I}}_{ij}^{-1} ,
\end{equation}
where the indices i and j denote the different parameters,
$\textrm{Cov}(\boldsymbol{\lambda})$ is the covariance of the 
parameters $\lambda$, and $\mathcal{I}_{ij}$ is defined as:

\begin{equation}
\label{eq:CRLB2}
\mathcal{I}_{ij}=\left\langle\frac{\partial \ln 
L(\boldsymbol{x},\boldsymbol{\lambda})}{\partial
\lambda_i}\frac{\partial \ln L(\boldsymbol{x},\boldsymbol{\lambda})}{\partial \lambda_j}
\right\rangle \equiv-\left\langle\frac{\partial^2
\ln L(\boldsymbol{x},\boldsymbol{\lambda})}{\partial
\lambda_i\partial
\lambda_j}\right\rangle
\end{equation} 

For Gaussian likelihood functions, 
like the pulsar timing likelihood function, 
$\mathcal{I}$ can be analytically calculated and the result is known as
the Slepian-Bangs formula \citep{sle1954,ban1971}, that is,

\begin{equation}
\label{eq:CRLB3}
\mathcal{I}_{ij}=\frac{\partial {\bf C}}{\partial \beta_i} {\bf C}^{-1}\frac{\partial 
{\bf C}}{\partial \beta_j}   + \frac{1}{2} {\rm  {\bf tr}}\Big\lbrack{\bf 
C}^{-1}\frac{\partial {\bf S(\lambda)^{\textrm{{\bf T}}}}}{\partial 
\lambda_i}{\bf C}^{-1} \frac{\partial {\bf S(\lambda)}}{\partial \lambda_j} 
\Big\rbrack\,.
\end{equation}

Here, $\beta_i$ are the model parameters describing the covariance matrix, $\lambda_i$, are the 
parameters describing the unknown waveform ${\bf S}$ and 
${\bf tr}$ is the matrix trace.
In the context of SSGW in this paper, we use the reduced likelihood 
function with the timing parameters marginalized \citep{vlm+2009},
where we now separate the timing residuals in the residuals induced
by stochastic processes, $\res_{\rm s}$ (for example red noise or a stochastic
GW background) and the deterministic residuals induced by the SSGW, $S(\lambda)$,
thereby $\res=\res_{\rm s}-S(\lambda)$.
In our specific problem, $\lambda$ is the set of seven parameters 
required to describe the monochromatic SSGW.
When focusing on the SSGWs, the 
terms with partial derivatives of {\bf C} are zero
and Eq.~\eqref{eq:CRLB3} reduces to

\begin{equation}
{\cal I}_{ij}=\frac{1}{2} {\rm  {\bf tr}}\Big\lbrack{\bf 
C}^{-1}\frac{\partial {\bf S(\lambda)^{\textrm{{\bf T}}}}}{\partial 
\lambda_i}{\bf C}^{-1} \frac{\partial {\bf S(\lambda)}}{\partial \lambda_j} 
\Big\rbrack\,.
	\label{eq:CRLB_single}
\end{equation}

\begin{figure}
\includegraphics[width=\columnwidth]{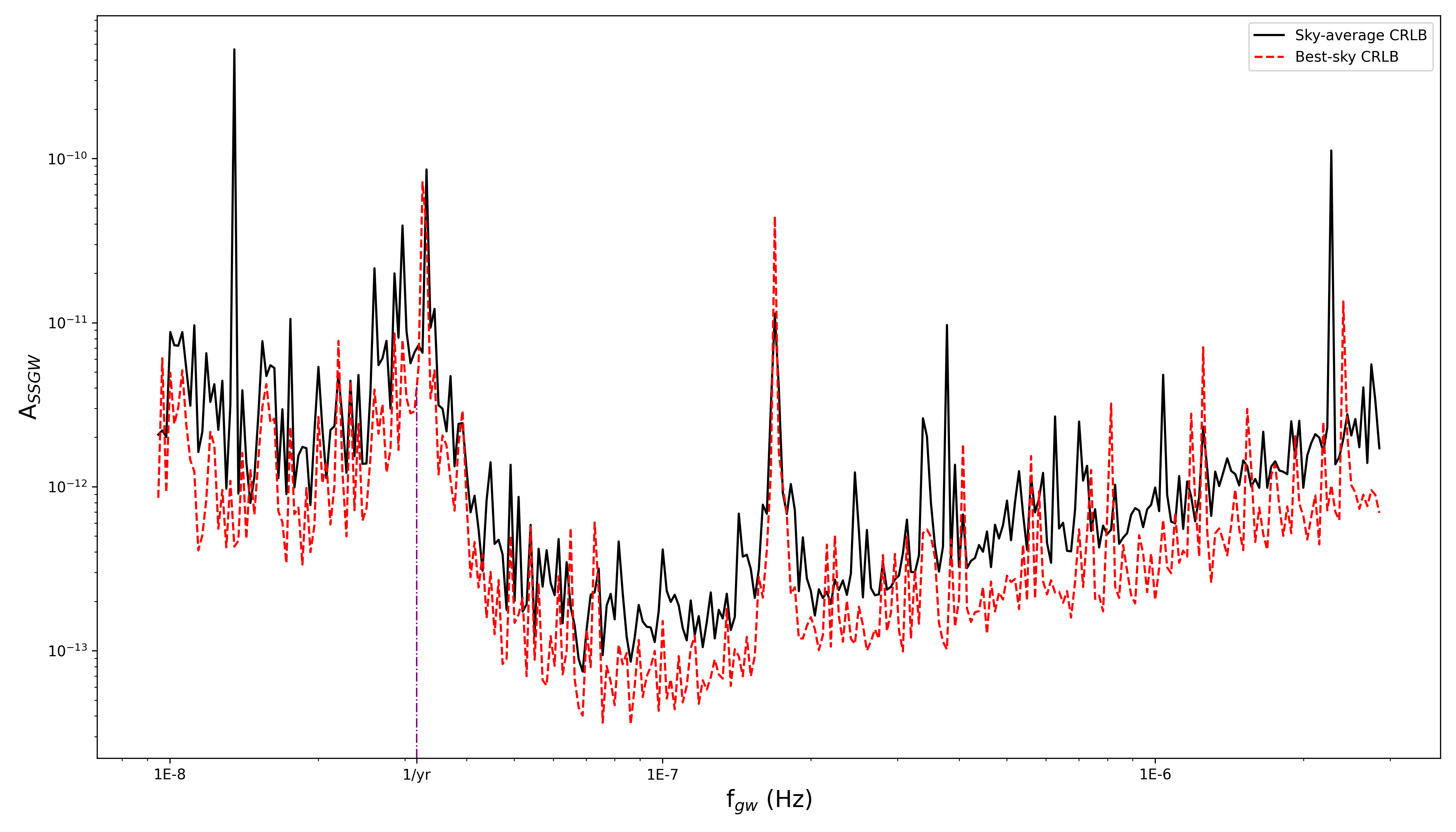}
\caption{The sky-average (solid, black line) and best-sky (dashed, red line) CRLBs for the SSGW strain amplitude
from GW-driven, circular SMBHBs.
Note that the CRLB corresponds to the 1-$\sigma$
error bar. The plot shows the 1/yr frequency with 
the dashed-dot, purple line. Note that due to the limited sensitivity
below 10$^{-7}$\,Hz, the 1/yr spike due to the Earth's orbital frequency
is not very narrow and distinct.}
\label{fig:GWcrlb}
\end{figure}

For the CRLB calculations,
the noise covariance matrix $\rm \bf C$, is fixed
using the maximum-likelihood values
of the noise components,
as derived from the pulsar noise analysis. 
We run 1000 analyses by selecting random values from 
uniform distributions for the 6 SSGW parameters 
(excluding the amplitude), 
and calculate the covariance of the amplitude
via Eq.~\ref{eq:CRLB1}. 
The sky-position-parameters range is set accordingly
depending whether we are interested in the full sky-average
amplitude limits or the `best-sky' limits.
Fig.~\ref{fig:GWcrlb} shows both results. Note that 
by default the estimated CRLB value corresponds to the
68\% confidence level (which corresponds to the 
$1-\sigma$ uncertainty, as we are dealing with a Gaussian
likelihood). From this result, we may expect the 
the sky-average and best-sky upper limits to have a factor of few
difference. More specifically, in the frequency range $10^{-7}-10^{-6}$Hz, 
the average difference is a factor of 3.35, 
while at frequency 1$\mu$Hz, the difference
is a factor of 2.2. 
We also note that for frequencies below $10^{-7}$, due to the limited
timespan, the sensitivity drops sharply and becomes non-comparable to 
previously published results. For this reason, we opted to only include
frequencies above $10^{-7}$ in the Bayesian analysis, 
saving computational resources and time.

\subsection{Bayesian Inference }
\label{sec:bayeslims}

Our main SSGW analysis follows
the Bayesian inference principles outlined
in Section~\ref{sec:bayes_theory}.
As we are only interested in amplitude upper limits,
we do not need to perform any model selection.
We perform two analyses.
First, as in the derivation of the CRLB,
we perform fix-noise analysis using 
the maximum-likelihood noise-parameter values.
This allows a direct comparison of the methods.
In addition, we perform a complementary analysis
with varying noise for five key GW frequencies,
namely 0.1, 0.3, 0.5, 1 and 2 $\mu$Hz.
This allows for the derivation of the most
conservative upper limit at our reference frequency
of 1$\mu$Hz.
We use uniform priors for the amplitude
with wide range
in order to derive robust upper limits.
The prior ranges are presented in Table~\ref{tab:bayes_settings}.

The sky-average results are derived by averaging 
the upper limits derived in 400 sky cells,
formed by having a 20$\times$20 grid in 
equally large in RA and cos(DEC). 
As explained earlier, motivated by the CRLB results, 
we limit our analysis in the GW frequency range of $10^{-7} - 3\times10^{-6}\,$Hz.
The amplitude upper limit is evaluated at each
GW frequency with a fixed frequency analysis,
using a frequency grid uniform in $\log_{10}$
for a total of 180 frequencies per sky cell.
The best-sky limit was calculated with one single analysis
where the RA and DEC priors were set 
at $\pm10^\circ$ around the 
pulsar position coordinates.

\begin{table} 
    \centering
    \caption{Ranges and types for the priors used in the 
    Bayesian SSGW upper-limit analysis. The prior types 
    Uni and log-Uni correspond to uniform priors in linear and logarithmic 
    space, respectively. The sky coordinates RA and $\cos$(DEC) are 
    marked with an asterisk ($^*$) to denote that the quoted range is
    the total covered, but each analysis used only the coordinates
    range corresponding to each sky cell. The GW angular frequency
    $\omega_{\rm gw}$, is fixed for each analysis,
    each time using a value from the pre-calculated grid.}
    \label{tab:bayes_settings}
    \begin{tabular}{c c c}
    \toprule
      Parameter & Priors & type \\
            \midrule             
        $A_{\textrm{SSGW}}$ & [$10^{-10}$,$10^{-20}$] & Uni\\
        RA$^*$ & [0,2$\uppi$] & Uni\\
        $\cos$(DEC)$^*$ & [1,-1] & Uni\\
        $\omega_{\rm gw}$ & [$10^{-7}$,3$\times10^{-6}$] & Uni/fixed from grid\\
        $\iota$ & [0,$\uppi$] & Uni\\
        $\psi$ & [0,2$\uppi$] & Uni\\
        $\phi_0$ & [0,2$\uppi$] & Uni\\
    \bottomrule
    \end{tabular}
\end{table}

\begin{figure}
\includegraphics[width=\columnwidth]{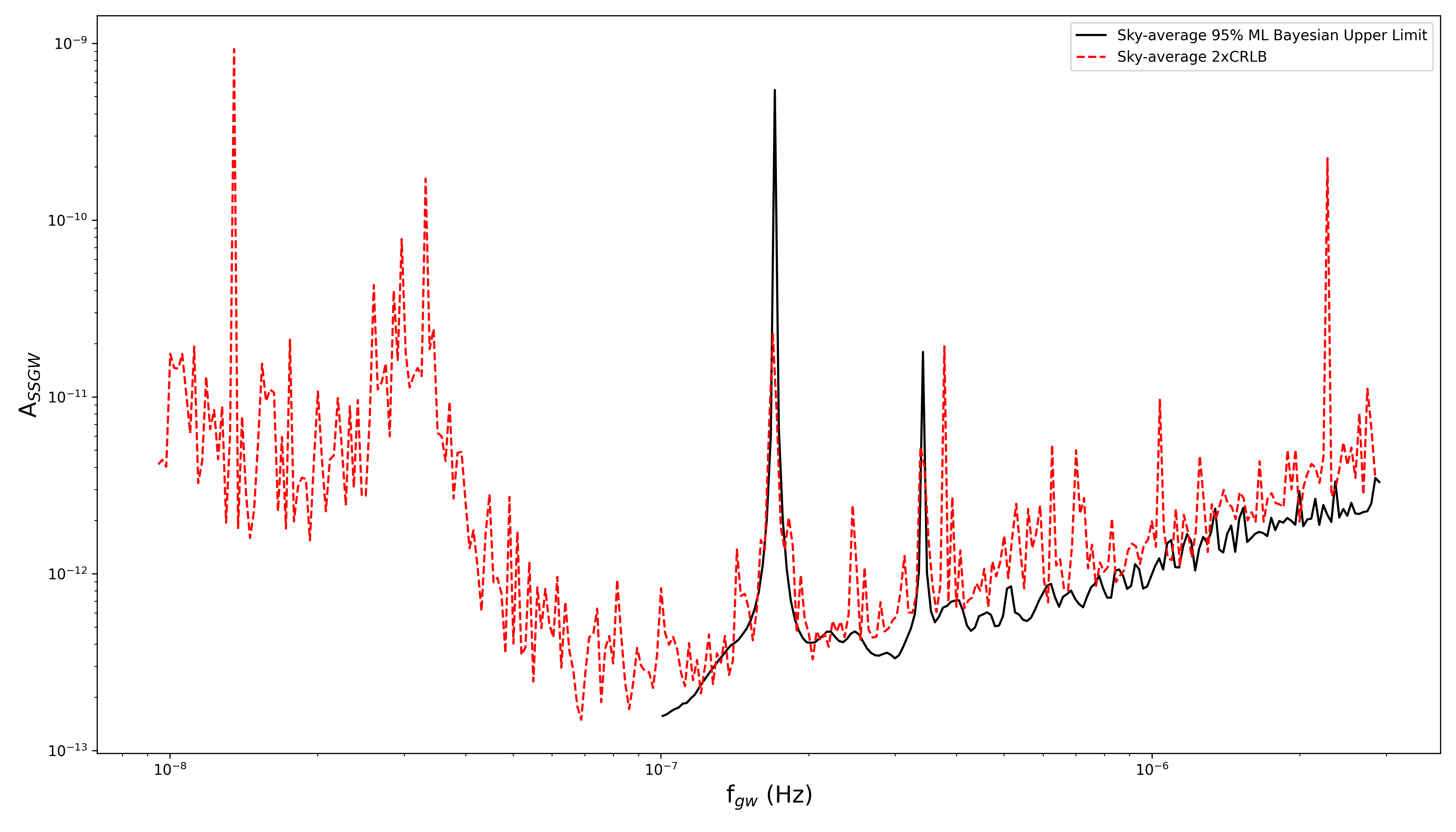}
\caption{Comparison between the Bayesian-derived upper limit (solid, black line) and the CRLB (dashed, red line) for the
SSGW strain amplitude from GW-driven, circular SMBHBs. 
Both analyses use fixed pulsar noise parameters, corresponding
to the maximum-likelihood (ML) from the single-pulsar noise analysis.
The Bayesian result corresponds to the 95\%  confidence level. We plot the 
double of the CRLB to approximate the 2-$\sigma$ error-bar, 
under the assumption of Gaussian parent distribution.}
\label{fig:GWbayesVcrlb}
\end{figure}

We first make a comparison between the Bayesian and the CRLB estimates.
Fig.~\ref{fig:GWbayesVcrlb} shows the two results.  As expected, the
results are compatible and in fact, almost identical, which suggests that
the Bayesian result approaches that 
of a fully efficient estimator \citep{fis63}.

\begin{figure}
\includegraphics[width=\columnwidth]{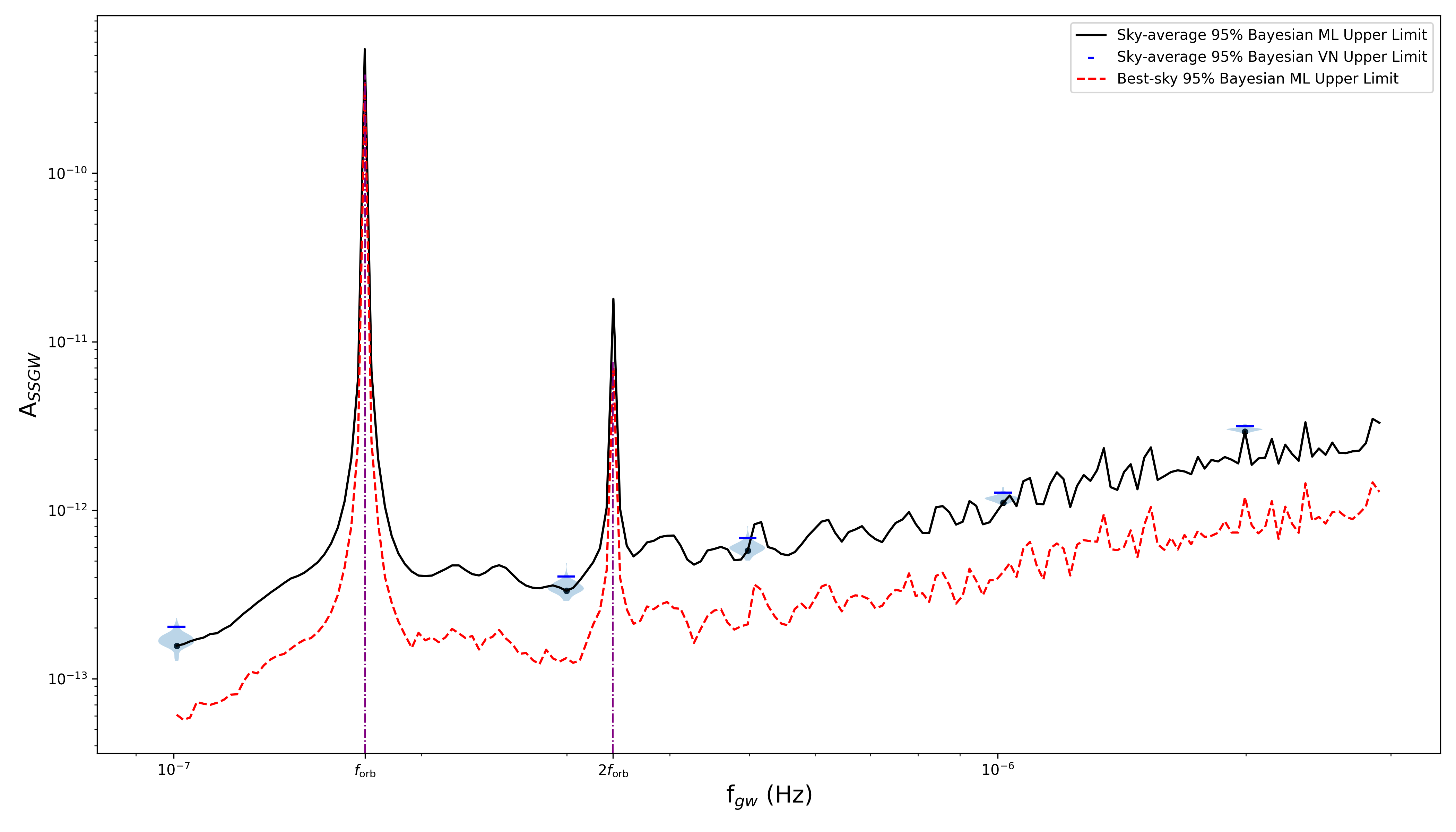}
\caption{Results from the Bayesian analyses:
The figure shows the sky-averaged fixed-pulsar noise analysis
using the maximum-likelihood (ML) noise parameters (solid, black line), 
and the same limit in the pulsar direction (best-sky position; dashed, red line). Results correspond to the 95\%  confidence level.
The blue violin plots show the posterior distribution
for the varying pulsar noise (VN) analysis performed for 5 GW frequencies.
The 95\% upper limits are marked with (blue) horizontal lines.
For better comparison, we also denote the 
corresponding upper limits on these frequencies
on the ML line, using filled circles. Finally, note that the two large spikes are
associated with the pulsar's orbital frequency and its first
harmonic (marked with dashed-dot purple lines), 
and neither should be confused with the 1/yr spike, which is not
in this figure's frequency range (see also note in Fig.~\ref{fig:GWcrlb})}
\label{fig:GWbayesfixVvaryVbestSky}
\end{figure}

Fig.~\ref{fig:GWbayesfixVvaryVbestSky} presents the results
from the Bayesian analyses. It compares the sky-average with the best-sky 95~\%
upper limits of the amplitude for the full 180-frequency spectrum
using fixed-noise analysis. 
In addition, we plot the results for the varying-noise, full-sky analysis for the 
five key frequencies, noting
the corresponding 95~\% amplitude upper limits.
At the reference GW frequency of 1\,$\mu$Hz
the most conservative sky-average upper limit, using
the varying pulsar-noise analysis, is 1.26$\times10^{-12}$.
The corresponding best-sky limit is 4.77$\times10^{-13}$.
For the fixed-noise analysis, the same sky-averaged and best-sky limits are
1.10$\times10^{-12}$ and 1.01$\times10^{-13}$, respectively.
In order to further highlight the effect of using a single pulsar, 
in which case the sensitivity to GWs tends to zero in the 
anti-pulsar direction, we note that the the sky-average amplitude limit
in half the sky of the pulsar direction is 7.01$\times10^{-13}$, 
in contrast to 1.41$\times10^{-10}$ in half the sky of the anti-pulsar direction.

\subsubsection{High-resolution sky map at 1$\mu$Hz}
\label{sec:map}

\begin{figure*}
\includegraphics[width=\textwidth]{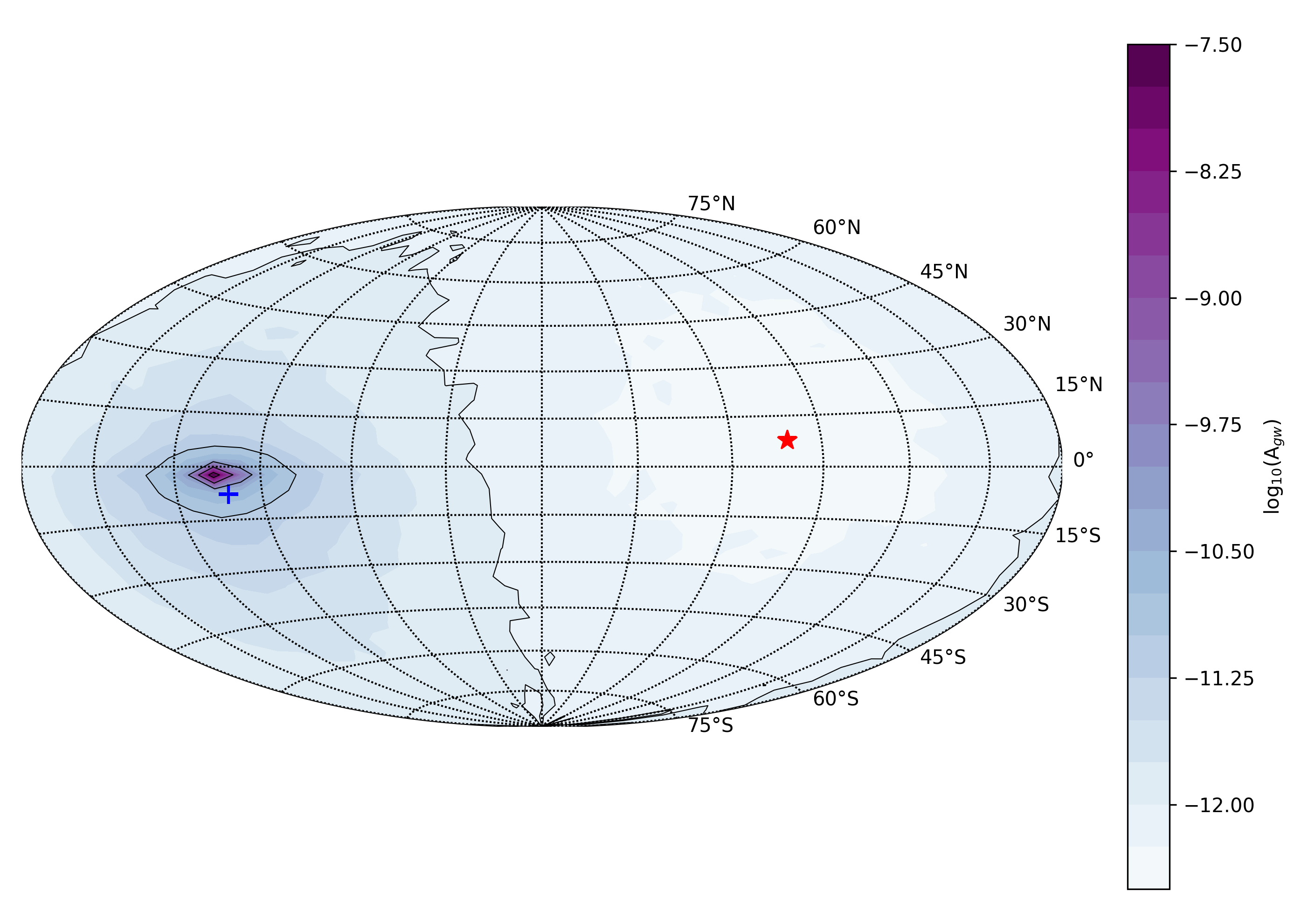}
\caption{Sky map of the logarithmic amplitude's upper limit (95\% confidence level) for SSGWs from circular SMBHBs, using the CPTA \psr{} data, at 1$\mu${}Hz GW frequency. The sky map is produced on the 40$\times$40 grid in RA and DEC. The red star and cross indicate the 
location of \psr{} and its opposite direction in the sky (anti-pulsar direction), respectively. The slight offset between the anti-pulsar direction and worst sensitivity is an artifact due to finite grid size.}
\label{fig:sky_map}
\end{figure*}

We repeat the Bayesian analysis on a finer, 40$\times$40
grid of 1600 sky cells at our reference GW frequency of 1$\mu$Hz
in order to create a high-resolution sky map of the 
SSGW strain amplitude upper limit. 
We use the fixed noise approach to reduce computational cost, 
as the numerical results are identical to those using the 20$\times$20,
and thus the sky map serves primarily for a visual representation
of the upper limit as a function of the sky location.
The sky map is presented in Fig.~\ref{fig:sky_map} and shows very clearly
the dipolar sensitivity of the single-pulsar GW detector:
the sensitivity maximizes in the pulsar direction and 
approaches zero at the opposite direction, which leads to
the measured multiple orders of magnitude difference in the sensitivity in their 
two respective sky hemispheres.

\section{Conclusions}
\label{sec:conclude}

We have used CPTA data collected with the FAST radio telescope from the
millisecond pulsar \psr{} to derive robust upper limits on the strain
amplitude of monochromatic SSGW from gravitational-wave driven SMBHBs
in circular orbits at the $\mu$Hz regime.  This is one of the most
observed MSPs in context of PTAs, due to its brightness that leads to
high-precision TOA measurements and low levels of time-correlated noise.
As it is observable from all PTAs both in the North and South hemisphere,
it has been a source used for various studies, including noise analysis at
very high frequencies, scintillation and pulse jitter studies.  So far,
it has provided the best SSGW strain amplitude limits at frequencies.
Due to their very high measurement precision, the CPTA FAST data are
especially suitable for SSGW studies, and we can probe the $\mu$Hz regime
thanks to dense-observations campaigns that give the dataset a best
cadence of 1-2 days.  Due to the recent profile-change event of \psr{}
in April 2021, the data beyond this date are not suitable for this study
and are excluded.
As the FAST observations began in mid-2019, our dataset has a timespan of only 1.67\,yr, which causes some limitations in the sensitivity to SSGW
across the spectrum, and especially to frequencies below $10^{-7}$\,Hz.
However, the high TOA precision offsets this loss and allowed us to
derive upper limits comparable to the previously published EPTA limits
\citep{psb+18} that used 4.3 years of data with roughly daily cadence.
Using Bayesian inference with the analysis software package \ftwo{}, we
derived a sky-average strain amplitude upper limit of 1.26$\times10^{-12}$
at 1\,$\mu$Hz
at 95\% confidence level, and a limit of 4.77$\times10^{-13}$ in the direction of \psr{}, where the sensitivity is maximum. We also produced a high-resolution sky map
of the amplitude limit by running the analysis on a fine grid of 1600 sky cells.

\begin{figure}
\includegraphics[width=\columnwidth]{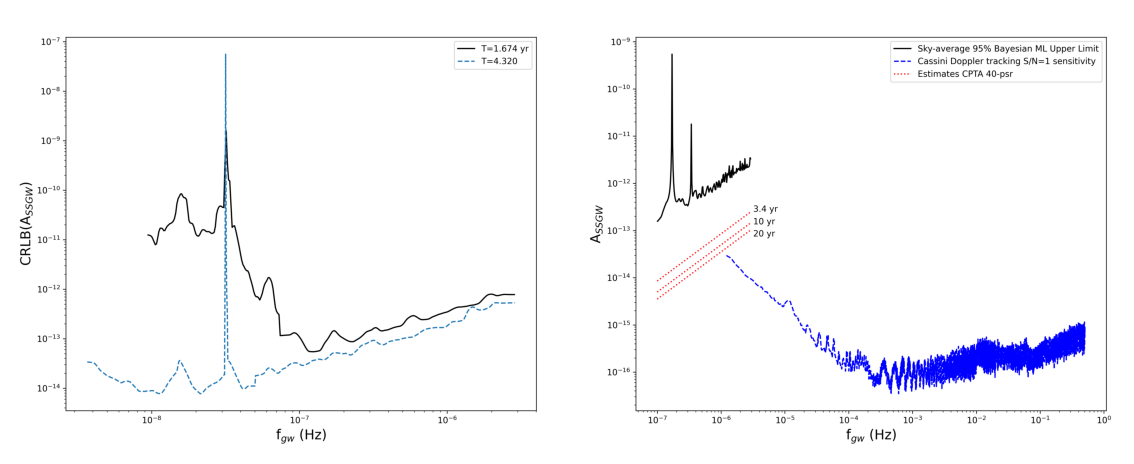}
\caption{Left panel: CRLB Sensitivity curves for \psr{} 
simulated data with a 2-day cadence for timespans of 1.67 (solid, black line) and 4.32\,yr (dashed, cyan line). The curves are smoothed for better visualization
of their difference.
The short timespan causes a sensitivity loss of 2.2 at the $\mu$Hz regime.
Right panel: 95\% confidence level sky-averaged upper limits for GW strain of \psr{} (solid, black line), and S/N=1 sensitivity of Cassini 2001 – 2002 GW observations (dashed, blue line), where the data comes from \cite{arm2006}. 
The red, dotted lines, show estimates for the CPTA, using 40 pulsars and
timespans of 3.4, 10 and 20\,yr.}
\label{fig:CPTAvDoppler}
\end{figure}

By comparison to the \cite{psb+18}, 
the sky-average sky limit is $\sim3.5$ times higher, 
and the best-sky limit $\sim2.2$ times higher.
This is primarily due to the shorter time-span in CPTA data, caused by the
profile change that limits the dataset. Figure~\ref{fig:CPTAvDoppler} (left panel)
shows the effect of the
short timespan using simulated \psr{} data. 
The two datasets have properties similar to the CPTA data, i.e.
ToA uncertainty of 50\,ns and a 2-day cadence. White noise includes
only radiometer noise to exclude jitter effects. The first dataset has the
timespan of the present CPTA data (1.67\,yr) and the second has
a timespan equal to the data in \cite{psb+18} (4.32\,yr).
Because of the short timespan and the 1/T limit near the 1/yr frequency spike, 
the shape of the sensitivity curve for the first case rises at 
such high frequency, that 
it affects the sensitivity in the $\mu$Hz regime. By comparison to the second
curve, the sky-average limit is 2.2 times higher. Note that this is the 
same difference with the best-sky limit when comparing the two real 
datasets, which satisfy the exact same condition ($\pm10^\circ$ around the pulsar position). The exact area around the opposite direction to the pulsar
excluded in sky-average result in \cite{psb+18} is not mentioned, 
and therefore we cannot have 
an exact comparison. Nevertheless, the simulations show that the majority
of the difference between the two results is explained by the short CPTA timespan.
Details on effects on sensitivity by other mechanisms,
such as jitter, is referred for future work.
We note however, that the majority of CPTA pulsar data are not jitter-, but
radiometer-noise limited.

In the current study, we only used data of a single pulsar, i.e. \psr{}, to prove the concept and verify the validity of the analysis software. Such a GW sensitivity is ten times worse comparing to the spacecraft tracking experiment \citep{arm2006} at the similar frequency band (see Fig.~\ref{fig:CPTAvDoppler}).
In future studies, data of multiple pulsars can be combined. The SSGW sensitivity increases with a rough scale of $N^{-1/2}$, where $N$ is the total number of data points. We expect future combination of CPTA data (50+ pulsars), will lower the current upper limit by about 10 times. 
Since most CPTA pulsar are radiometer-noise limited, we compute 
a first estimate on the improvement in the SSGW upper limits
we can achieve when including the rest of the CPTA data in the analysis,
without being concerned for jitter effects.
In Fig.~\ref{fig:CPTAvDoppler} (right panel), we show an estimate of the upper limit
if we assume the use of 40 pulsars with timing precision at the average 
of the 40 brighter CPTA sources (Xu et al., in prep.). We assume timespans of 3.4 (timespan of
CPTA first data release), 10 and 20\,yr. 
For simplicity, we assume the pulsars to be isotropically distributed in the sky.
We assume that all pulsars except \psr{} can contribute beyond the 1.67-yr limitation,
and therefore counteract the corresponding sensitivity loss of $\sim2$ discussed above.
One can expect more improvement over time
with dense observing campaigns for various pulsrs
as the one made for \psr{} 
and the addition of more pulsars in
PTA target lists.
The longer data span will also further aid us in probing the lower frequency band, i.e. $f\le 10^{-8}$. We can thus expect an upper limit at the level of $\lesssim 10^{-14}$ in the near future for the SSGW, 
resulting in pulsar timing probably becoming be the most sensitive $\mu$Hz GW detector.

\normalem

\begin{acknowledgements}

Observation of CPTA is supported by the FAST Key project. FAST is a Chinese national mega-science facility, operated by National Astronomical Observatories, Chinese Academy of Sciences. This work is supported by the National SKA Program of China (2020SKA0120100), the National Nature Science Foundation grant no. 12041303 and 12250410246, the CAS-MPG LEGACY project, and funding from the Max-Planck Partner Group. KJL acknowledges support from the XPLORER PRIZE and 20-year long-term support from Dr. Guojun Qiao. HX is supported by Major Science and Technology Program of Xinjiang Uygur Autonomous Region No. 2022A03013-4. The data analysis are performed with computer clusters \textsc{DIRAC} and \textsc{C*-system} of PSR@pku and computational resource provide by the \textsc{PARATERA} company. 

HX, SYC, YJG, JJC, BJW, JWX, ZHX, RNC, YHX, and KJL are core team to perform the data analysis for the current paper, where HX worked on data reduction, timing and data analysis; RNC and ZHX performed the noise and SSGW analysis; BJW carried out single pulse studies; JJC and JWX calibrated data polarization; ZHX and RNC searched for GW single sources; YHX studied scintillation processes.

\end{acknowledgements}

\bibliographystyle{raa}
\bibliography{ms2024-0082.bib}

\end{document}